\documentclass[prd,showpacs,twocolumn,aps]{revtex4-1}
\usepackage{slashed}
\usepackage{mathrsfs}
\usepackage{amsfonts}
\usepackage{amsmath}
\usepackage{amssymb}
\usepackage{revsymb}
\usepackage{graphicx}
\usepackage{mathrsfs}
\usepackage{bm}
\usepackage{bbm}
\usepackage{psfrag}
\usepackage{color}
\usepackage{hyperref}
\usepackage{natbib}
\usepackage[usenames,dvipsnames]{xcolor}

\newcommand{\erf}{\trm{erf}}

\newcommand{\be}{\begin{equation}} 
\newcommand{\ee}{\end{equation}} 
\newcommand{\bea}{\begin{eqnarray}} 
\newcommand{\eea}{\end{eqnarray}} 

\newcommand{\eps}{\varepsilon}
\newcommand{\mbf}[1]{\mathbf{#1}}
\newcommand{\trm}[1]{\textrm{#1}}

\newcommand{\figref}[1]{Fig. \ref{#1}}

\newcommand{\eqnref}[1]{Eq. (\ref{#1})}
\newcommand{\eqnrefs}[2]{Eqs. (\ref{#1}) and (\ref{#2})}

\newcommand{\Ecr}{E_{\trm{cr}}}

\newcommand{\vphi}{\varphi}

\newcommand{\defto}{=}

\newcommand{\xmp}{x^{-\,\prime}}
\newcommand{\xpl}{x^{+}}
\newcommand{\xmi}{x^{-}}

\newcommand{\es}{\pmb{\eps}_{s}}
\newcommand{\ep}{\pmb{\eps}_{p}}
\newcommand{\ev}{\pmb{\eps}}

\newcommand{\ks}{\widehat{\mbf{k}}_{s}}
\newcommand{\kp}{\widehat{\mbf{k}}_{p}}

\newcommand{\LHE}{\mathcal{L}_{\trm{HE}}}

\newcommand{\ms}{\zeta}
\newcommand{\ampp}{\mathcal{E}_{p}}
\newcommand{\amps}{\mathcal{E}_{s}}

\newcommand{\colr}[1]{#1}

\begin{document}
\title{Interaction of photons traversing a slowly varying electromagnetic 
background}
\author{B. \surname{King}}
\email{ben.king@plymouth.ac.uk}
\affiliation{
    School of Computing and Mathematics, Plymouth University, 
    Plymouth PL4 8AA, UK\\
    Arnold Sommerfeld Center for Theoretical Physics, \\ 
Ludwig-Maximilians-Universit\"at M\"unchen,
    Theresienstra\ss e 37, 80333 M\"unchen, Germany}

\author{P. \surname{B\"ohl}}
\email{patrick.boehl@physik.uni-muenchen.de}
\affiliation{Arnold Sommerfeld Center for Theoretical Physics, \\ 
Ludwig-Maximilians-Universit\"at M\"unchen,
    Theresienstra\ss e 37, 80333 M\"unchen, Germany}
    
\author{H. \surname{Ruhl}}
 \email{hartmut.ruhl@physik.uni-muenchen.de}
\affiliation{Arnold Sommerfeld Center for Theoretical Physics, \\ 
Ludwig-Maximilians-Universit\"at M\"unchen,
    Theresienstra\ss e 37, 80333 M\"unchen, Germany}

\date{\today}
\begin{abstract}
When two electromagnetic fields counterpropagate, they are modified 
due to mutual interaction via the polarised virtual electron-positron states of 
the vacuum. By studying how photon-photon scattering effects such as 
birefringence and four-wave mixing evolve as the fields pass through one 
another, we find a significant increase during overlap \colr{when both electromagnetic 
variants can be non-zero}. The results have particular relevance for calculations based on a 
constant 
field background.
\end{abstract}
%\pacs{12.20.-m 42.50.Ct 52.27.Ep 52.65.-y }
\pacs{}
\maketitle

%%%%%%%%%%%%%%%%%%%%%%%%%%%%%%%%%%%%%%%%%%%%%%%%%%%%%%%%%%%
\section{Introduction}
That electromagnetic fields can polarise virtual 
electron-positron pairs of the vacuum has been known since the early pioneering 
calculations of Sauter \cite{sauter31}, Halpern \cite{halpern34}, Weisskopf 
\cite{weisskopf36} and Heisenberg and Euler \cite{heisenberg_euler36}, later 
being rederived in the language of quantum electrodynamics by Schwinger 
\cite{schwinger51}. The polarised pairs facilitate the process of 
photon-photon scattering, which can be broadly split into inelastic processes 
such as vacuum pair-creation and elastic processes where fermion states do not 
persist on the mass shell. There are many predicted manifestations of elastic 
effects. The polarisation of scattered photons could be used to verify this 
phenomenon through
birefringence, polarisation rotation \cite{baier67b, narozhny69, adler71, 
baier75a, 
dipiazza06, heinzl06, king10b} and helicity flipping \cite{baier75a,dinu13}. 
The propagation direction of scattered photons could also be used and signals of
diffraction \cite{king10a, tommasini10, 
monden11, king12} and reflection \cite{hatsagortsyan11, gies13, dinu14}
have been calculated. Also in the frequency of scattered photons, signals can 
occur through the process of four-wave mixing \cite{lundstroem_PRL_06, king12}, 
photon-splitting \cite{bialynicka-birula70, adler71, affleck87, brodin07, 
dipiazza07} and  
photon-merging \cite{bb81, dipiazza05, fedotov07, gies14}.
\newline

These phenomena are of interest in astrophysics, for example to describe 
the behaviour of magnetised neutron 
stars \cite{baring91,harding97,baring01,ho03,harding06,gitman12,costa13,adorno14b}, particularly in
astrophysical electromagnetic shocks \cite{fabrikant82,heyl98,heyl99} and in high-intensity laser 
physics \cite*{marklund_review06, 
ehlotzky09, dipiazza12}, being searched for in terrestrial experiments 
\cite{pvlas12, rizzo12, uggerhoj13}.
\newline

When photon wavelengths are much shorter than the length on which pair 
creation occurs, photon-photon scattering can be 
described using an effective theory for interacting 
electromagnetic fields given by the
 Heisenberg-Euler Lagrangian.
Typically one considers the effect on some weak 
``probe'' field, which can be a single photon, as it passes through a
``strong'' field. In applications to potential laser experiments, it is the 
asymptotic state of the probe field which is of primary interest as detection 
apparatus is necessarily far removed from the interaction region. In 
simulations of astrophysics in the magnetospheres of neutron 
stars, one typically calculates the effect on 
propagating photons in a classical magnetic field, which is taken to vary 
adiabatically, with the constant-field solution being integrated over 
macroscopic regions in kinetic equations \cite{baring91,thompson95}.
\newline

In the current paper, we focus on the evolution of an oscillating 
probe field that scatters in a slowly-varying strong background, with both 
fields being described as plane waves. \colr{We will often refer to an ``overlap'' of fields, which 
is equivalent to the largest amplitude of the two electromagnetic invariants, defined in the 
following section}. Using the 
Heisenberg-Euler Lagrangian, we 
will identify a signal of elastic photon-photon scattering that increases with the 
overlap of the fields and disappears when the overlap tends to zero. 
Moreover, we will find that this scattered ``overlap field'' can be much larger 
than the ``asymptotic'' scattered field which persists after the probe has 
passed through the background, particularly for parameters considered in 
high-intensity laser experiments. The presence of the overlap field implies a 
difference in the predicted physics when one calculates effects in a 
forever-constant background compared to \colr{those} in a constant background
evolved adiabatically from the infinite past. Furthermore, the overlap field is 
neglected whenever an approximation to elastic photon scattering in 
inhomogeneous fields is made by integrating over forever-constant 
background scattering rates.
%%%%%%%%%%%%%%%%%%%%%%%%%%%%%%%%%%%%%%%%%%%%%%%%%%%%%%%%%%%

%%%%%%%%%%%%%%%%%%%%%%%%%%%%%%%%%%%%%%%%%%%%%%%%%%%%%%%%%%%
\section{Analytical method}
\begin{figure}[!h]
\noindent\centering
\hspace{0.025\linewidth}
\includegraphics[draft=false, width=0.42\linewidth]{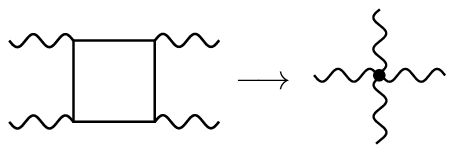}
\hfill\includegraphics[draft=false, 
width=0.42\linewidth]{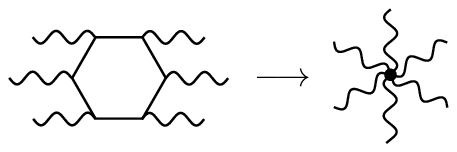}\hspace{0.025\linewidth}
\caption{The Heisenberg-Euler Lagrangian contains effective vertices for 
classical electromagnetic fields interacting via the quantum effects 
described by the four-photon scattering box diagram and six-photon scattering 
hexagon diagram.}
\label{fig:46diag} 
\end{figure}

Let us consider the electromagnetic field to be the sum of a weak 
probe and strong background field
\bea
F^{\mu\nu} = F_{p}^{\mu\nu}+F_{s}^{\mu\nu},
\eea 
where $F$ is the Faraday tensor \cite{jackson75} and the subscripts $_{p}$ and 
$_{s}$ pertain, throughout the paper, to the probe and strong fields respectively. If 
one defines dimensionless electromagnetic and secular invariants,
\bea
\mathcal{F} = -F^{2}/4\Ecr^{2}, &\quad& \mathcal{G} =
-FF^{\ast}/4\Ecr^{2},\label{eqn:fieldinvariants}\\
a = \left[\sqrt{\mathcal{F}^{2}+\mathcal{G}^{2}}+\mathcal{F}\right]^{1/2}, 
&\quad& b =
\left[\sqrt{\mathcal{F}^{2}+\mathcal{G}^{2}}-\mathcal{F}\right]^{1/2}\!\!,
\eea
where $F^{2}=F^{\mu\nu}F_{\mu\nu}$, $FF^{\ast}=F^{\mu\nu}F^{\ast}_{\mu\nu}$, 
giving $\mathcal{G} = \mathbf{E}\cdot\mathbf{B}$ and 
$\mathcal{F}=(E^{2}-B^{2})/2$, in which $E^{2}=\mbf{E}\cdot\mbf{E}$ and 
electric 
and magnetic fields $\mbf{E}$, $\mbf{B}$ are dimensionless, having been 
normalised by the critical field strength $\Ecr=m^{2}/e$. We set here 
and throughout $\hbar=c=1$. \colr{The one-loop effective action in a constant external field is 
given by the Heisenberg-Euler Lagrangian} \cite{schwinger51}
\bea
\mathcal{L}_{\trm{HE}} &=& -\frac{\alpha 
m^{4}}{8\pi^{2}}\int_{0}^{\infty}\!\!ds\,\frac{\mbox{e}^{-s}}{s^{3}}
\Big[s^{2}ab\,\trm{cot}\,as\,\trm{coth}\,bs - 1 \nonumber\\
&&\qquad\qquad\qquad\qquad+\frac{s^{2}}{3}(a^{2}-b^{2})\Big].\label{eqn:LEHfull}
\eea
As we are interested in the effects on electromagnetic fields and wish to avoid a discussion on 
pair creation, we perform a weak-field expansion of \eqnref{eqn:LEHfull} for when $E\ll1$:
\bea
\mathcal{L}_{\trm{HE}} &=& \frac{m^{4}}{\alpha}\sum_{n=1}^{\infty} 
\mathcal{L}_{n},\label{eqn:LEHwf}\\
\mathcal{L}_{1} &=& 
\frac{\mu_{1}}{4\pi}\left[\left(E^{2}-B^{2}\right)^{2} + 
7(\mbf{E}\cdot\mbf{B})^{2}\right]\label{eqn:L1}\nonumber\\
\mathcal{L}_{2} &=& 
\frac{\mu_{2}}{4\pi}\left(E^{2}-B^{2}\right)\left[2\left(E^{2}-B^{2}
\right)^{2}
+13\left(\mbf{E}\cdot\mbf{B}\right)^{2} \right
],\nonumber\label{eqn:L2}\\
\mathcal{L}_{3} &=& 
\frac{\mu_{3}}{4\pi}\left[3\left(E^{2}-B^{2}\right)^{4}
+22\left(E^{2}-B^{2}\right)^{2}\left(\mbf{E}\cdot\mbf{B}\right)^{2}\right. 
\nonumber\\
&&\qquad +\left. 19\left(\mbf{E}\cdot\mbf{B}\right)^{4} \right
],\label{eqn:L3}
\eea
where $\mu_{1}=\alpha/90\pi$, $\mu_{2}=\alpha/315\pi$, 
$\mu_{3}=4\alpha/945\pi$ (the fine-structure constant occurs in the 
denominator in the Lagrange densities \eqnref{eqn:L3} due to rewriting 
fields in terms of the critical field). 
The term $\mathcal{L}_{n}$ describes the effective scattering of $2n$ photons 
and we will restrict ourselves to the leading-order effects of 
$\mathcal{L}_{1}$ corresponding to effective 
four-photon scattering or the ``box diagram'', and $\mathcal{L}_{2}$ corresponding to effective 
six-photon scattering or the ``hexagon diagram'', as demonstrated in \figref{fig:46diag}. \colr{It 
has been shown that in the low-frequency limit $\omega/m \ll 1$, a direct calculation of 
four-photon scattering agrees with the leading-order term in the above weak-field expansion 
\cite{karplus50}. 
A further and more restrictive condition can be placed on the frequencies we consider, when we 
demand that the work performed by the external field over the reduced Compton wavelength is less 
than the electron rest energy $\omega/mE\ll1$.}
Applying the Euler-Lagrange equations to 
$\mathcal{L}=\mathcal{L}_{\trm{MW}}+\mathcal{L}_{\trm{HE}}$, where
$\mathcal{L}_{\trm{MW}}=m^{4}(E^{2}-B^{2})/8\pi\alpha$ leads to the classical 
Maxwell equations, one arrives at a wave equation modified by vacuum 
polarisation:
\bea
\square\,\mbf{E} = \mbf{T}[\mbf{E},\mbf{B}],
\eea
where we have defined a source term:
\bea
\mbf{T} = 4\pi\left[\nabla \wedge \partial_{t}\mbf{M} + 
\partial_{t}^{2}\mbf{P}-\nabla\left(\nabla\cdot\mbf{P}\right)\right]
\label{eqn:Tfull}
\eea
for \colr{dimensionless} magnetisation 
$\mbf{M}=(\alpha/m^{4})\partial\LHE/\partial\mbf{B}$ and 
polarisation $\mbf{P}=(\alpha/m^{4})\partial\LHE/\partial\mbf{E}$. To simplify the 
discussion, let us consider the probe and strong fields to 
counterpropagate with normalised wavevectors $\widehat{\mbf{k}}_{p}=(0,0,1)$, 
$\widehat{\mbf{k}}_{s}=(0,0,-1)$ and calculate scattering along the axis of 
symmetry. This effectively reduces the system to one spatial and one 
temporal dimension. \colr{An interesting consequence of this is that the charge density, given by 
$\nabla\cdot\mbf{P}$, disappears. This is due to the electromagnetic field having no component in 
the direction of inhomogeneity, which is the direction of propagation along the $z$-axis. 
Therefore, the final term in \eqnref{eqn:Tfull} can be neglected.} Assuming the change in the 
fields due to 
scattering is small, we then solve:
\bea
\left(\partial_{t}^{2}-\partial_{z}^{2}\right)\mbf{E} = 
\mbf{T}[\mbf{E}^{(0)}], \label{eqn:we2}
\eea
where 
$\square\,\mbf{E}^{(0)}=\square\,\mbf{B}^{(0)}=0$ 
are vacuum solutions to the wave equation, and 
$\mbf{B}^{(0)}_{j}=\widehat{\mbf{k}}_{j}\wedge\mbf{E}^{(0)}_{j}$ for 
$j\in\{s,p\}$. \colr{In particular, we will choose $\mbf{E}^{(0)}(x^{-},x^{+}) = 
\mbf{E}_{p}(x^{-})+\mbf{E}_{s}(x^{+})$, where $x^{\pm}=t\pm z$}. 
\colr{We wish to solve the scattering problem for when two initially well-separated excitations of 
the electromagnetic field $\mbf{E}_{p}(x^{-})$ and 
$\mbf{E}_{s}(x^{+})$ that vanish on the boundary ($\lim_{x^{\pm}\to\pm\infty}\mbf{E}_{p,s}=0$) 
collide at some finite $t$ and $z$.} The 
solution to \eqnref{eqn:we2} is acquired using:
\bea
\mbf{E}(t,z) &=& 
\mbf{E}^{(0)}(t,z) + 
\Delta\mbf{E}(t,z)\label{eqn:Eform1}\\
\Delta\mbf{E}(t,z) &=&  
\int\!dt'\,dz'\,G_{\trm{R}}(t-t',z-z')\mbf{T}\left[\mbf{
E}^{(0)}(t',z')\right],\nonumber\label{eqn:DelE}\\
\eea
where $G_{\trm{R}}$ is the retarded Green's function for the wave equation 
in one spatial and one temporal dimension \cite{mahan02}
\bea
G_{\trm{R}}(t,z)= \frac{1}{2}\theta(t)\theta\left(\frac{t}{v}-|z|\right),
\eea
for propagation speed $v$, which we will assume to be equal to the speed of 
light $v=1$ in all calculations in this paper. Applying this method to 
\eqnref{eqn:we2}, we have
\bea
\left(\partial_{t}^{2}-\partial_{z}^{2}\right)\Delta\mbf{E} = 
\mbf{T}\left[\mbf{E}^{(0)}\right]. \label{eqn:we3}
\eea
\colr{The approximation $\mbf{T}[\mbf{E}]\approx \mbf{T}[\mbf{E}^{(0)}]$ in 
\eqnrefs{eqn:DelE}{eqn:we3} can be understood by the following argument.
Since the source $\mbf{T}$ contains, to leading order in $E\ll1$, the cube of electromagnetic 
fields $E^{3}$, the lowest order 
neglected term is $\sim  E^{2}\,\Delta E$. An approximation for $\Delta E$ can be made by 
using \eqnref{eqn:DelE}, which turns out to give $\Delta E\sim \alpha E^{3}L_{\vphi}$ for some 
phase length $L_{\vphi}$. Therefore, if $\alpha E^{2} L_{\vphi}\ll1$, the approximation of 
neglecting the vacuum's influence on the driving fields when calculating vacuum polarisation, 
$\mbf{T}[\mbf{E}]\approx \mbf{T}[\mbf{E}^{(0)}]$, can be justified. We note the importance not only 
of field strength, but also of phase length.}
\newline

Through partial integration in $t$, the 
scattered field becomes a sum of 
forward- (positive $z$-direction) and backward-propagating scattered 
fields
\bea
\Delta\mbf{E}(t,z) &=&  
\Delta\overrightarrow{\mbf{E}}(t,z) 
+\Delta\overleftarrow{\mbf{E}}(t,z), 
\label{eqn:delep}
\eea
where boundary terms can be neglected when the initial overlap of fields is 
zero, and
\bea
\Delta\overrightarrow{\mbf{E}}(t,z) &=& 
\int_{-\infty}^{z}\!\frac{dz'}{2}~\mbf{J}(
\xmi+z',z')\label{eqn:DelEforward}\\
\Delta\overleftarrow{\mbf{E}}(t,z) &=& \int_ { z } ^ { 
\infty}\!\frac{ 
dz'}{2}~\mbf{J}(\xpl-z',z'),\label{eqn:DelEbackward}
\eea
where $x^{\pm}=t\pm z$ and $\mbf{J}$ is the current occurring in 
Maxwell's equations
\bea
\mbf{J}(t,z)=4\pi\left[\widehat{\mbf{k}}_{p} \wedge \partial_{z}\mbf{M}(t,z) + 
\partial_{t}\mbf{P}(t,z)\right]. \label{eqn:J}
\eea
\colr{The interpretation of $\Delta\overrightarrow{\mbf{E}}(t,z)$, 
$\Delta\overleftarrow{\mbf{E}}(t,z)$ as the forward- and backward-scattered field respectively can 
be seen more clearly by rewriting \eqnrefs{eqn:DelEforward}{eqn:DelEbackward} in lightcone 
co-ordinates (where the substitution $y=2(z'-z)$ has been made):
\bea
\Delta\overrightarrow{\mbf{E}}(\xmi,\xpl) &=& 
\int_{-\infty}^{0}\!\frac{dy}{4}~\mbf{J}(
\xmi,\xpl+y)\label{eqn:DelEforwardLC}\\
\Delta\overleftarrow{\mbf{E}}(\xmi,\xpl) &=& \int_ { 0 } ^ { 
\infty}\!\frac{ 
dy}{4}~\mbf{J}(\xmi-y,\xpl).\label{eqn:DelEbackwardLC}
\eea
Therefore $\Delta\overrightarrow{\mbf{E}}(t,z)$ remains constant on the probe-field lightcone, 
($x^{-}$ constant i.e. forward-scattered) and 
$\Delta\overleftarrow{\mbf{E}}(t,z)$ on the strong field lightcone ($x^{+}$ constant i.e. 
backward-scattered).}
 
\subsection{Overlap and asymptotic field}
To make clearer what is happening, we calculate by way of example, 
part of the forward-scattered field \eqnref{eqn:DelEforward} arising from 
the second term in the current \eqnref{eqn:J} using \eqnrefs{eqn:Eform1}{eqn:DelE}. The polarisation 
is
\bea
\mbf{P}[\mbf{E}] = \frac{\mu_{1}}{2\pi}\left[2\left(E^{2}-B^{2}\right)\,\mbf{E}+7 
\left(\mbf{E}\cdot\mbf{B}\right)\,\mbf{B}\right],
\eea
where we recall we consider $\mbf{P}[\mbf{E}^{(0)}]$ and since 
$\mbf{E}^{(0)}(t,z)=\mbf{E}_{s}(x^{+})+\mbf{E}_{p}(x^{-})$ and 
similarly for the magnetic field, which are both plane waves, we see different 
combinations of powers of $E_{s}$ and $E_{p}$ will occur in $\mbf{P}$. For 
brevity, let us focus on terms proportional to the probe field squared. Then the 
corresponding part of the scattered field is
\bea
\pmb{\eps}  \int_{-\infty}^{z}\!\frac{dz'}{2}
\partial_{t'}\left(E_{p}^{2}E_{s}\right)
\eea
where $\pmb{\eps}$ is the polarisation vector that absorbs all other constants 
in this example, \colr{$E_{p}=E_{p}(t'-z')$, $E_{s}=E_{s}(t'+z')$} and the derivative is 
evaluated at 
$t'=x^{-}+z'$, which becomes
\bea
-\pmb{\eps}  
\int_{-\infty}^{z}\!\frac{dz'}{2}
\partial_{z'}\left(E_{p}^{2}\right) E_{s}
+\pmb{\eps}  
\int_{-\infty}^{z}\!\frac{dz'}{2}
E_{p}^{2}\partial_{z'}\left(E_{s}\right). \label{eqn:asysplit1}
\eea
\eqnref{eqn:asysplit1} is the \emph{asymptotic} plus the \emph{overlap} 
field respectively. To elaborate these labels, we can use that the 
derivatives are evaluated on the lightcone of the probe field so that 
\eqnref{eqn:asysplit1} becomes:
\begin{gather}
 \frac{1}{2}\pmb{\eps} E_{p}^{\prime}(\xmi)E_{p}(\xmi)
 \int_{-\infty}^{0}\!dy~
 E_{s}(\xpl+y)
 \nonumber \\
 +\frac{1}{2}\pmb{\eps}\,E_{p}^{2}(\xmi)  
 E_{s}\left(\xpl\right),
 \label{eqn:asysplit2}
\end{gather}
where $^{\prime}$ indicates the derivative. For the 
first 
term, we see that on the probe lightcone (e.g. $\xmi=0$), long after the 
collision in the asymptotic limit $t,z\to\infty$, the term remains (assuming 
the integration over the strong field is non-vanishing). Therefore we 
label this the \emph{asymptotic} scattered field. The second term 
corresponds to a surface term and the strong and probe fields are evaluated on 
their respective light cones. When the overlap of the fields, \colr{or equivalently the amplitude 
of the field invariants}, tends to zero, so 
does this term and therefore we label this the \emph{overlap}
scattered field. We note that if a constant field is adiabatically evolved 
from the infinite past, the overlap field \emph{is} generated. This should be 
contrasted with the case of an ever-present constant field, in which the 
overlap field vanishes identically. 
\newline

In this example we considered the probe field squared, corresponding to 
generation of a second harmonic (the frequency of the strong field is taken to 
be much smaller than that of the probe), also referred to as 
``photon-merging''. A division into overlap and asymptotic scattered fields can be made in each 
combination of powers of strong and probe fields that 
occur in the interaction.
\newline

To investigate these ideas, we will choose the probe 
and strong fields in this paper to be of the form
\bea
\mbf{E}_{p}(\varphi_{p}) &=& 
\ep\,\ampp\,\mbox{e}^{-\left(\frac{\varphi_{p}}{\Phi_{p}}
\right)^ { 2 } }
\cos\varphi_{p}\\
\mbf{E}_{s}(\varphi_{s}) &=& 
\es\,\amps\,\mbox{e}^{-\left(\frac{\varphi_{s}}{\Phi_{s}}\right)^{2}},
\eea
where $\Phi_{j}=\omega_{j}\tau_{j}$ and 
$\varphi_{j}=k_{j}^{\mu}x_{\mu}~$ for $j\in\{s,p\}$,
$\ep\cdot\ep=\es\cdot\es=1$ and we are interested in the case 
$\omega_{p}\tau_{s}\gg 1$.
%%%%%%%%%%%%%%%%%%%%%%%%%%%%%%%%%%%%%%%%%%%%%%%%%%%%%%%%%%%

%%%%%%%%%%%%%%%%%%%%%%%%%%%%%%%%%%%%%%%%%%%%%%%%%%%%%%%%%%%
\section{Numerical method}
The numerical solution of the nonlinear Maxwell equations following from 
the sum of classical and Heisenberg-Euler Lagrangians is based on
the PCMOL (Pseudo Characteristic Method of Lines) \cite{carver1980pseudo}, 
matrix inversion and the CVODE ODE-Solver from SUNDIALS (SUite of Nonlinear 
and 
DIfferential/ALgebraic equation Solvers) 
\cite{Hindmarsh:2005:SSN:1089014.1089020}.
Since we assume propagation only in the $z$-direction and only transverse
polarisations, the resulting equations of motion can be written in
matrix form:
  \bea  
\left(\mathbbm{1}_{4}+\mbf{A}\right)\partial_{t}\mbf{f}+\left(\mathbf{Q}+\mbf{B}
\right)\partial_{z}\mbf{f}=0,\label{eqn:MatrixForm1D}
  \eea
  where $\mbf{f} = (E_{x},E_{y},B_{x},B_{y})^{T}$, 
$\mathbbm{1}_{4}=\trm{diag}(1,1,1,1)$
  is the identity matrix in four dimensions,
  $\mbf{Q}=\mathrm{adiag}(1,-1,-1,1)$ is an anti-diagonal matrix and
  $\mbf{A}\defto (a_{ij})$ and $\mbf{B}\defto (b_{ij})$ are the
  nonlinear corrections resulting from \eqnref{eqn:LEHwf}
with $a_{ij}=b_{ij} =
  0$ for $i>2$. For the weak field expansion $\mathcal{L}_1$ ($\mathcal{L}_2$), 
the components are
  quadratic (fourth order) polynomials of the field
  components. 
  \newline
  
  Let us first consider the linear case with
  $\mbf{A}=\mbf{B} =0$. In the PCMOL, one uses the diagonalisability of the 
matrix $\mbf{Q}$, which means one can find a basis 
$\mbf{u}:=\mbf{P}\mbf{f}$ 
such 
that $\mbf{\Lambda} \defto \mbf{P} 
\mbf{Q}\mbf{P}^{-1}=\mathrm{diag}(-1,-1,1,1)$ is 
diagonal with real eigenvalues:
   \begin{align}
   \mbf{P}\!=\!\frac{1}{\sqrt{2}}
     \left( \begin{smallmatrix}
      \text{-}1&   0& 0& 1\\
        0&   1& 1& 0\\
       1&   0& 0& 1\\
        0& \text{-}1& 1& 0
     \end{smallmatrix} \right)
     \quad
     \mbf{u}\! :=\! 
\mbf{P}\mbf{f}=\!\frac{1}{\sqrt{2}}\left(\begin{smallmatrix}
       B_y-E_x\\ E_y+B_x\\ E_x+B_y\\ B_x-E_y 
     \end{smallmatrix}\right)\label{eq:Ubasis}
     \end{align}
     The new set of equations is given by:
       \begin{equation*}
   \partial_{t}\mbf{u}+\mbf{\Lambda}\,\partial_{z}\mbf{u}=0. 
\label{eqn:MatrixFormLinear1D}
   \end{equation*}
  The eigenvalues are called the characteristic speeds and the
  positive (negative) sign corresponds to a component propagating in
  the positive (negative) $z-$direction. The system, which is taken to
  be of a length of $320\,\mu\trm{m}$, is 
discretised in
  space using $N=2\cdot10^{5}$ points. The four-dimensional vector
  $\mbf{u}$ can then be mapped onto a $4N$-dimensional one, 
$\mbf{u}= (\ldots 
u_4^{i-1}u^i_1u^i_2u_3^iu_4^iu_1^{i+1}\ldots)$, where $0<i\leq N$ labels the 
grid point. The spatial derivatives of 
the components
  $u^i_j$ are 
approximated with upwind-biased finite differences
  determined by the sign of the characteristic speed. This is done
  using fourth-order stencils \cite{schiesser1991numerical}, where the
  values of the derivative near the boundary are also approximated
  with fourth-order accuracy using grid points only inside the simulation box.
Since the
  derivatives at one point are calculated with the field values at the
  specific and surrounding points, the action of the derivative can be
  written as a matrix multiplication: $\partial_z \mbf{u} \approx
  \mbf{D} \mbf{u}$, where $\mbf{D}$ is a $4N\times4N$ matrix. In the PCMOL, the 
equations are now transformed back to the original basis $\mbf{f}$, but the 
system is solved in $\mbf{u}$, which is completely equivalent. This has the 
advantage of 
automatically implementing open boundary conditions due to the upwind character 
in the single components. The electric and magnetic fields are then obtained by 
applying 
$\mbf{P}^{-1}$ for output at each grid point.
\newline

  We now consider the nonlinear case. To bring the system to an
  ODE form $\mbf{u}'(t)=f(\mbf{u},t)$ ($f$ is called the ``right-hand-side 
function'', the $'$ denotes the time derivative), we need to invert the matrix
  $(\mathbbm{1}_4+\mbf{A})$. Since $\mbf{A}$ is a local operator of the field
  components, it is only necessary to consider the inversion for each single 
grid
  point. We rewrite $\mbf{A}$ as  $\mbf{A}=\mbf{M}\mbf{N}$ with
\begin{equation}
  \mbf{M}=
 \begin{pmatrix}
    1 & 0 \\
    0 & 1 \\
    0 & 0\\
    0& 0
  \end{pmatrix}, \quad \quad\mbf{N} = 
  \begin{pmatrix}
     a_{11} & a_{12} & a_{13} & a_{14}\\
   a_{21} & a_{22} & a_{23} & a_{24}\\
\end{pmatrix}.
\end{equation}
and apply the Woodbury Formula \cite{golub2012matrix},
\begin{align}
 (\mathbbm{1}_4+\mathbf{A})^{-1} = \mathbbm{1}_4 - \mbf{M}(\mathbbm{1}_2+ 
\mbf{NM})^{-1}\mbf{N}\quad \label{eq:Woodbury}
\end{align}
to reduce the inversion of the $4\times 4$-matrix $(\mathbbm{1}_4+\mbf{A})$ 
to one of the $2\times2$ matrix
\begin{align}
  (\mathbbm{1}_2 +\mbf{N}\mbf{M}) = \begin{pmatrix}
    1+a_{11} & a_{12} \\
    a_{21}  & 1+a_{22}
    \end{pmatrix},
  \end{align}
    which is performed at each evaluation of the
  right-hand-side function $f$ via an 
LU-factorisation. Since our method employs a weak 
field expansion, we expect only small
corrections from the nonlinearities $\mbf{A}$ and $\mbf{B}$, such that
the the matrix
  $(\mathbbm{1}_4 + \mbf{A})^{-1}(\mbf{Q}+\mbf{B})$ has similar spectral 
properties (i.e. the same signs of the eigenvalues) as
$\mbf{Q}$. Therefore we use the same biased differencing as in the linear case.
The system is now solved using the parallel, extended-precision version of 
CVODE, where the nonlinear right-hand-side function
  is given by
  \begin{align}
  f(\mbf{u},t) =-\mbf{P}(\mathbbm{1}_4 + 
\mbf{A})^{-1}(\mbf{Q}+\mbf{B})\mbf{P}^{-1} \mbf{D}\mbf{u}.
\end{align}
$\mbf{P}(\mathbbm{1}_4 +
\mbf{A})^{-1}(\mbf{Q}+\mbf{B})\mbf{P}^{-1}$ is now a block-diagonal $4N\times 
4N$ matrix with $4\times 4$ 
blocks
acting on each grid point, as explained above.
We use the provided Adams-Moulton methods and the functional iteration to solve 
the corresponding linear system of equations.
The signals are analysed using a spatial Fourier Transform in \textit{Wolfram 
Mathematica} \cite{mathematica9} and the frequency components are filtered 
under the assumption $\omega = |\mbf{k}|$ and transformed back to spatial 
co-ordinates. To analyse the DC-component, we subtract the analytical
expression of the strong pulse from the signal. 
%%%%%%%%%%%%%%%%%%%%%%%%%%%%%%%%%%%%%%%%%%%%%%%%%%%%%%%%%%%

%%%%%%%%%%%%%%%%%%%%%%%%%%%%%%%%%%%%%%%%%%%%%%%%%%%%%%%%%%%
\section{Competing vacuum processes}
For clarity, we consider each frequency component of the scattered field 
separately and neglect the change in frequency due to the background frequency 
scale $1/\tau_{s} \ll \omega_{p}$. To leading order in $\amps,\ampp\ll1$, the
scattered field can be written as
\bea
\Delta\mbf{E} &=& 
\sum_{l=1}^{\infty} 
\ampp^{l}\mbox{e}^{-l\left(\frac{\varphi_{p}}{\Phi_{p}}\right)^{2}}\left[
\mbf{C}_{l}\sin l\varphi_{p}+\widetilde{\mbf{C}}_{l}\cos l\varphi_{p}\right ] 
\nonumber \\
&& \qquad + \ampp^{2}\left[\mbf{C}_{0}+\widetilde{\mbf{C}}_{0}\right] 
\label{eqn:delE2}.
\eea
In \eqnref{eqn:delE2} we note that the scattered field is written as a sum over
harmonics of the probe field. Each higher harmonic involves a higher 
power of 
$\ampp \ll 1$ so in general higher-harmonics are less likely in this 
regime. For each harmonic we then note two spacetime-dependent vector terms 
with coefficients $\mbf{C}_{l}$ and $\widetilde{\mbf{C}}_{l}$.
The $\mbf{C}_{l}$ terms are out of phase with the probe field and form 
the asymptotic field whereas the $\widetilde{\mbf{C}}_{l}$ terms are in phase 
with the probe field and correspond to the overlap field. Although we neglect 
processes of a higher order than four- and six- photon scattering in the current 
analysis, they can be calculated straightforwardly using the method used here. \colr{We highlight 
the fact that 
$\lim_{\amps\to0}\mbf{C}_{l}=\lim_{\amps\to0}\widetilde{\mbf{C}}_{l}=\lim_{\amps\to0}\mbf
{C}_{0}=\lim_{\amps\to0}\widetilde{\mbf{C}}_{0}=0$, showing that $\Delta\mbf{E}$ vanishes in the 
limit 
where the strong or probe field is absent.} In the following we comment on the first few harmonics.

\subsection{Fundamental harmonic}
If the scattered field is much weaker than the probe, it can be described 
by analogy with a modified refractive index, $1+\delta n$, $\delta n \ll 1$. 
The 
probe field lightcone then becomes $\varphi_{p}=\omega_{p}[t/(1+\delta n) 
-z]$. Expanding $\cos\varphi_{p}$ in $\delta n$, the leading-order scattered 
field is in antiphase with the probe field, so this effect should be 
entirely covered by the asymptotic field in our analysis. For the current 
scenario we find:
\bea
\mbf{C}_{1} &=& -\mu_{1} 
\ev_{s,1}\amps^{2}\,\frac{\omega_{p}\tau_{s}\sqrt{\pi}}{
\sqrt{2}}\frac{1+\erf\,(\sqrt { 2 } \varphi_{s}/\Phi_{s})}{2}\label{eqn:C1}\\
\widetilde{\mbf{C}}_{1} &=& -\mu_{1} 
\ev_{s,1}E_{s}^{2}(\varphi_{s}),\label{eqn:Ct1}
\eea
where the polarisation of the scattered field is given by 
\bea
\pmb{\eps}_{s,1} &=& 
c_{1,1}~\es +c_{1,2}~\ks\wedge\es\label{eqn:eps1s}
\eea
\colr{with coefficients
\bea
c_{1,1}&=&4\,\es\cdot\ep\,(1-\ks\cdot\kp),\\ 
c_{1,2}&=&7\,(\es\cdot\kp\wedge\ep+\ep\cdot\ks\wedge\es).
\eea 
In particular, we notice that when $\kp=\ks$, $c_{1,1}=c_{1,2}=0$ and vacuum polarisation effects 
disappear, as they must in a single plane wave background \cite{schwinger51}}. \colr{The 
polarisation vector of the scattered field in all harmonics will be a function of these 
coefficients, so we highlight that $c_{1,1}$ originates from evaluating 
$\mathcal{F}_{ps}=-F^{\mu\nu}_{p}F_{s\,\mu\nu}/4\Ecr^{2}$ and $c_{1,2}$ from evaluating 
$\mathcal{G}_{ps}=-F^{\mu\nu}_{p}F^{\ast}_{s\,\mu\nu}/4\Ecr^{2}$. Therefore, a considerable 
simplification occurs when $\es\parallel\ep$ implying $c_{1,2}\to0$ or when $\es\perp\ep$ implying 
$c_{1,1}=0$. A consistency check of \eqnrefs{eqn:C1}{eqn:Ct1} can be performed by calculating the 
implied altered dispersion relation for the probe field.} We note that the 
well-known modified refractive index for $\kp = -\ks$ in a constant background is given by 
\cite{baier67b}:
\bea
\delta n(\varphi_{s}) = 
\frac{2\alpha\,E_{s}^{2}(\varphi_{s})}{45\pi}\left[4\left(\pmb{\eps}_{p}
\cdot\pmb { \eps } _ { s } \right)^ { 2 } +7\left(\pmb{
\eps}_{p}\wedge\pmb{\eps}_{s}\right)^{2}\right]. \label{eqn:dnconst}
\eea
If the corresponding phase difference $\delta\varphi_{p}$ is 
calculated by integrating \eqnref{eqn:dnconst} over the shape of 
$E_{s}$ in the following way:\colr{
\bea
\delta\varphi_{p}(z') = \omega_{p}\int^{z'}_{-\infty}\,dz~\delta 
n(\varphi_{s})\Big|_{t=z-\xmp},
\label{eqn:dphi}
\eea}
then the asymptotic field and \eqnref{eqn:C1} can be recovered exactly.

The presence of the overlap field in the fundamental harmonic cannot be 
described by a modified index of refraction. If the background is wider than 
several probe wavelengths ($\omega_{p}\tau_{s}\gg 1$), then the 
amplitude of the overlap field is much smaller than of the asymptotic in the fundamental harmonic. 
Both parts of the scattered field have the same polarisation as the probe in this case.

\subsection{Second harmonic}
\begin{figure}[!h]
\noindent\centering
\includegraphics[width=0.99\linewidth]{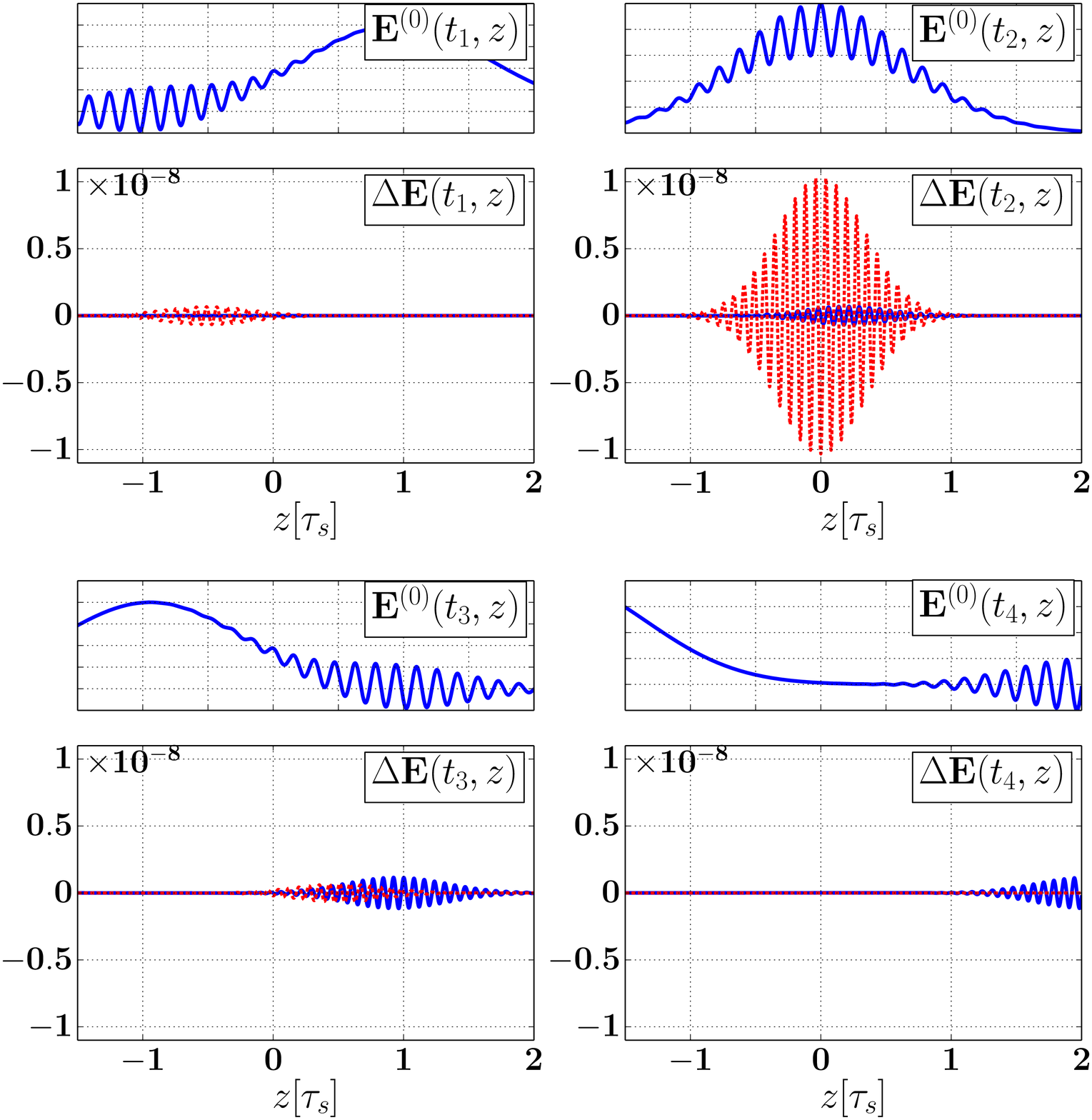}
\caption{The smaller panels plot \colr{snapshots} of the total electric field 
above the larger panels showing the corresponding state of the scattered overlap (red dashed) and 
asymptotic (blue solid) second harmonic field at times $t_{4}>t_{3}>t_{2}>t_{1}$. The 
electric fields are in units of the probe field amplitude, $\ampp$.}
\label{fig:4panel} 
\end{figure}
\begin{figure}[!h]
\noindent\centering
\includegraphics[width=0.8\linewidth]{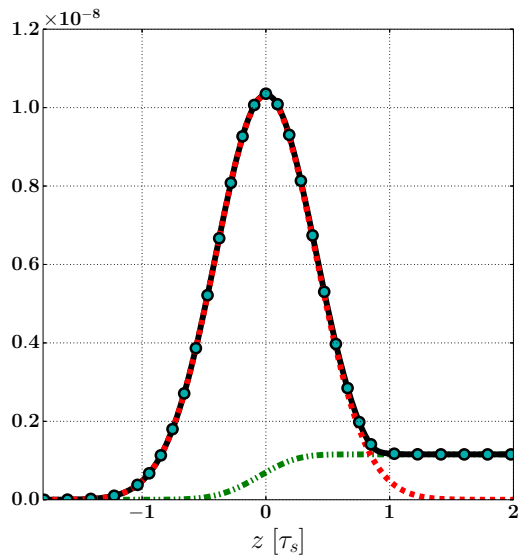}
\caption{The second harmonic overlap field (dashed line) generated in 
four-photon scattering can dominate the asymptotic field (dot-dashed 
line) generated in six-photon scattering. Agreement is also shown between 
simulation (points) and theory 
(solid line) for $\amps=0.02$, $\ampp=0.005$, $\omega_p = 
0.6\,\trm{eV}$, $\tau_s= 6.4\lambda_{p}$ and $\tau_p=5\lambda_{p}$, where the 
field is in units of the probe amplitude, $\ampp$.}
\label{fig:resplot} 
\end{figure}

The strongest contribution to the scattered field with double the frequency of 
the probe originates from four- and six-photon scattering. We find
\bea
\mbf{C}_{2} &=& -\mu_{2}\, 
\ev_{s,2}\amps^{3}\,\frac{\omega_{p}\tau_{s}\sqrt{\pi}}{
\sqrt{3}}\frac{ 1+\erf (\sqrt { 3 } \varphi_{s}/\Phi_{s})}{2}\label{eqn:C2}\\
\widetilde{\mbf{C}}_{2} &=& -\frac{\mu_{1}}{2} 
\ev_{p,1}E_{s}(\varphi_{s})-\frac{\mu_{2}}{2} 
\ev_{s,2}E^{3}_{s}(\varphi_{s})\label{eqn:Ct2}
\eea
where 
\bea
\pmb{\eps}_{p,1} &=& 
c_{1,1}~\ep +c_{1,2}~\kp\wedge\ep\label{eqn:eps1p}\\
\pmb{\eps}_{s,2} &=& 
c_{2,1}~\es +c_{2,2}~\ks\wedge\es\label{eqn:eps2s},
\eea
with $c_{2,1}=3\,c_{1,1}^{2}/2+13\,c_{1,2}^{2}/49$ and 
$c_{2,2}=13\,c_{1,1}c_{1,2}/14$. In one temporal and 
one spatial dimension, merging of two-photons via four-photon scattering in a strictly constant 
background 
is suppressed for kinematical reasons. However, when the background 
contains some inhomogeneity, the second harmonic \emph{can} be generated. This 
is also the case when a constant background is adiabatically evolved from the 
infinite past. Since the second-harmonic overlap field is of order $\alpha^{2}$ 
and the asymptotic field of order $\alpha^{3}$, there is a range of parameters 
for which the overlap field dominates. Let us define the gauge- and 
relativistically- invariant parameter $\ms$:
\bea
\ms = 
\int_{-\infty}^{\infty}\!d\varphi_{s}~\ms(\varphi_{s})
\eea
where $\ms(\varphi_{s}) = [\chi(\varphi_{s})]^{2}/\eta$, $\chi = 
\sqrt{|k_{p}F_{s}|^{2}}/m$ is the so-called 
quantum non-linearity parameter \cite{ritus85} and $\eta = k_{p}k_{s}/m^{2}$. 
For the current scenario, $\ms=\amps^{2}\omega_{p}\tau_{s}\sqrt{2\pi}$ and by 
comparing \eqnrefs{eqn:C2}{eqn:Ct2}, we note that when $\ms \ll 1$, the overlap 
field can dominate. The evolution of the scattered field is illustrated in 
\figref{fig:4panel}, and directly compared with the position of the probe and 
strong fields. In \figref{fig:resplot}, the maximum of the amplitude of the 
simulated second-harmonic signal is plotted and the evolution for the 
asymptotic and overlap fields compared. In the second harmonic, the rate of 
change 
of the overlap field is proportional to the gradient of the background. In 
\figref{fig:resplot} we observe that the maximum of the 
amplitude of the overlap field initially increases to an overall 
maximum when the probe and strong fields most overlap, after which the second 
harmonic is further generated field but phase-shifted by $\pi$ and 
destructively interferes with the already present second harmonic field. 

\subsection{Higher harmonics}
In the presence of some background inhomogeneity, a given harmonic is 
generated in the overlap field at one order in $\alpha$ lower than in the 
scattered field. For 
example, in one spatial and one temporal dimension, no asymptotic third harmonic signal is 
generated from the hexagon diagram, but an overlap signal \emph{is} permitted. 
We find 
\bea
\widetilde{\mbf{C}}_{3} = -\frac{\mu_{2}}{4}
\ev_{p,2}E^{2}_{s}(\varphi_{s}),
\eea
where $\ev_{p,2} = 
c_{2,1}\,\,\ep +c_{2,2}\,\,\kp\wedge\ep$
and the leading order term in $\mbf{C}_{3}$ can be found by calculating octagon 
diagram in the weak-field Heisenberg-Euler expansion. 
\newline

Although the overlap and asymptotic fields have different spacetime 
dependencies, we find that the polarisation selection rules for higher 
harmonic generation are identical. In particular
\begin{gather}
l\gamma_{\parallel} \to \gamma^{\prime}_{\parallel}; \qquad 2l\gamma_{\perp} 
\to \gamma^{\prime}_{\parallel }\qquad (2l-1)\gamma_{\perp} \to 
\gamma^{\prime}_{\perp} \label{eqn:polchannels}
\end{gather}
where $\gamma_{\parallel}$ corresponds to a probe photon obeying 
$\ep\wedge\es=\mbf{0}$, $\gamma_{\perp}$ to $\ep\cdot\es=0$, 
$l\in\mathbb{N}^{+}$ and $\gamma'$ is the photon generated through scattering. 
Therefore odd harmonics exhibit a slightly different 
polarisation 
behaviour and in particular admit a photon-merging cascade in the $\perp$ 
component. However, since this requires a minimum of three photons to merge, 
it is 
presumably only of relevance when the probe photon density is very high or 
path length very long. 
Another feature of this mechanism is that probe photons that are in a 
superposition of linear polarisations \emph{can} access the 
$\gamma_{\perp}+\gamma_{\perp}\to\gamma'_{\perp}$ channel, but only once. This 
can be seen by the coefficient of the outgoing $\perp$ channel depending on the 
overlap of probe photon $\perp$ and $\parallel$ components (e.g. $c_{2,2}$ in 
\eqnref{eqn:eps2s}). Once scattered, the merged photons are then confined to 
residing in a polarisation eigenmode 
thereafter.
\newline

The lowest-order non-trivial effect of the polarised vacuum on probe 
photons is 
a modification of the index of refraction (\eqnref{eqn:dnconst}) leading to
$k^{2} \neq 0$. We note that taking this non-trivial dispersion into 
account, more harmonics can be generated for a given-order diagram 
in which photons are no longer described by null fields as $F^{2}\neq 0$. 
In particular, the signal must no longer be in harmonics 
of the incoming field.  We will postpone analysis of this particular problem, 
which requires longer path lengths, for a future publication.

\subsection{Zeroth harmonic}
\begin{figure}[!h]
\noindent\centering
\includegraphics[width=0.8\linewidth]{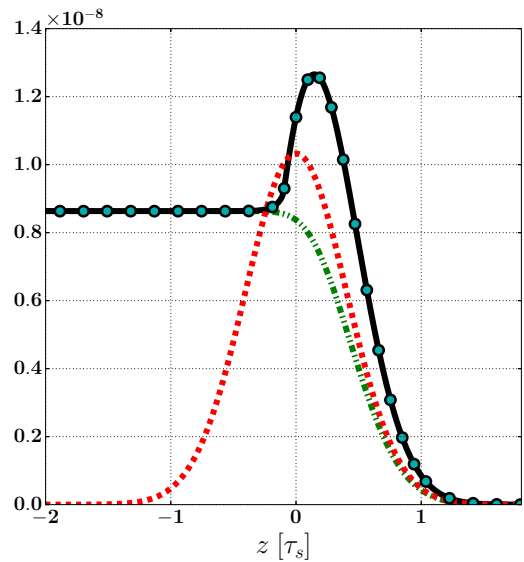}
\caption{Both 
asymptotic (dot-dashed line) and overlap (dashed line) signals for 
frequency down-conversion originate 
from four-photon scattering. The leading-order asymptotic signal is due 
to the change in background due to interaction with the probe. Theory (solid 
line) and simulation (points) agree and show a backwards-propagating signal. The
parameters are the same as in \figref{fig:resplot} and fields in units of the 
probe amplitude, $\ampp$.}
\label{fig:DCplot} 
\end{figure}

With the zeroth harmonic, DC component, or rectification, we are referring to a 
signal with the low frequency $\approx 1/\tau_{s} \ll 
\omega_{p}$ of the background. One probe photon is absorbed by and one photon 
emitted from the polarised vacuum pairs, leaving a photon of the frequency 
associated with the background. From momentum conservation, the scattered field 
has a momentum vector in the backwards direction. We find
\bea
\mbf{C}_{0} &=& -\mu_{1}\pmb{\eps}_{p,1}\amps  
\sqrt{
\frac { \pi } { 2 
}}\frac{\tau_{p}}{\tau_{s}}\frac{\varphi_{s}}{\Phi_{s}}\mbox{e}^{
-\left(\frac { \varphi_{s} }
{\Phi_{s}}\right)^{2}}\,\frac{1+\erf\left(\sqrt{2}\varphi_{p}/\Phi_{p}
\right)}{2}\nonumber\\ 
 \widetilde{\mbf{C}}_{0} &=& 
 -\frac{1}{2}\, 
\mu_{1}\pmb{\eps}_{p,1}\,E_{s}\left(\varphi_{s}\right)\,\mbox{e}^{-2\,
\left(\frac{\varphi_{p} }{\Phi_{p}}\right)^{2}} .
\eea
Emission in the backwards-direction is demonstrated in 
\figref{fig:DCplot}, and contrasts with the photon-merging behaviour shown in 
\figref{fig:resplot}.
Moreover, when the background varies, both asymptotic and overlap signal 
are generated in four-photon scattering. The polarisation of the DC 
component is then parallel to the strong field $\gamma_{\parallel}$.
\newline

Frequency down-conversion can also produce non-DC components, for 
example in $\gamma+\gamma\to \gamma + \gamma'$, for probe photons $\gamma$, 
the scattered photon $\gamma'$ is at the fundamental 
frequency. However, the more photons that participate, the 
smaller the effect when $\ampp\ll 1$. We stress the difference of frequency
down-conversion from photon splitting, as in the current case, no photon 
quantum is being split into quanta of lower energy. 
%%%%%%%%%%%%%%%%%%%%%%%%%%%%%%%%%%%%%%%%%%%%%%%%%%%%%%%%%%%

%%%%%%%%%%%%%%%%%%%%%%%%%%%%%%%%%%%%%%%%%%%%%%%%%%%%%%%%%%%
\section{Discussion}
Our results suggest that when one approximates photon-photon 
scattering in spacetime-varying fields by assuming that scattering is at each 
instant equivalent to \colr{that} in a strictly constant background, 
important physics 
is missed. We have seen how this even arises when a photon propagates through 
a quasi-constant background, due to the photon-photon interaction current 
involving derivatives of combinations of fields. When this current is 
integrated over, part of the scattered field is generated by a surface term 
that depends on the state of the background in the photon's past. Therefore it 
can occur that changes in the background strength, even when over 
distances much larger than the photon wavelength, can lead to 
a significant contribution to the rate of photon-photon scattering. In 
particular, the predicted evolution of the total scattered field is different 
for a strictly constant background compared to a constant background that has 
been adiabatically evolved from the infinite past. This could have potential 
implications for effects calculated in the overlap of probe and strong fields 
that rely on the instantaneously-constant background approximation such as the 
so-called ``vacuum resonance'' 
\cite{ho01,ho04} in strongly-magnetised pulsars, which could be searched for 
in a programme similar to the GEMS mission \cite{ghosh13}.
\newline

The intense magnetic field of certain neutron stars offers an excellent 
possibility to study strong-field quantum electrodynamical effects using
polarisation measurements of emitted photons \cite{ghosh13,taverna13}. The 
process of 
photon-splitting has been hypothesised to be of particular importance in 
the magnetospheres of neutron stars. Here we compare the density of photons 
(number per unit volume) that split $\rho_{\gamma\to2\gamma'}$ with those 
that merge
$\rho_{2\gamma\to\gamma'}$ in a quasi-constant magnetic field of 
strength $B$. We repeated the calculation leading 
to \eqnrefs{eqn:C2}{eqn:Ct2}, first taking the limit 
$\tau_{p,s}\to\infty$ and setting the 
background electric field to zero but allowing
a background field strength difference $\Delta B$ over the seed 
photons' history, defining $\Delta = \Delta B/B$. We then find:
\bea
\widetilde{\rho}_{2\gamma\to\gamma'} &=& 2\alpha^{3} 
\left[\frac{11\pm3}{180\pi}\right]^{2} 
B^{2}\Delta^{2}\frac{\omega}{m} (\rho\pi\lambdabar^{3})~\rho \\
\rho_{2\gamma\to\gamma'} &=& 8\alpha^{3} \left[\frac{37\mp 
11}{315\pi}\right]^{2} 
B^{2}\ms^{2}\frac{\omega}{m} (\rho\pi\lambdabar^{3})~\rho\\
\rho_{\gamma\to2\gamma'}&=& 
 \frac{\alpha^{3}}{10}\left[\frac{19}{315\pi}\right]^{2} 
 \frac{L}{\lambdabar} B^{6} 
 \left(\frac{\omega}{m}\right)^{5}\rho, 
\eea
where $\widetilde{\rho}_{2\gamma\to\gamma'}$ and $\rho_{2\gamma\to\gamma'}$ refer to merging 
in the overlap and asymptotic fields respectively, $\rho$ is the density of seed photons with 
frequency $\omega$, $L$ 
is propagation distance of the seed photon, $\zeta=B^{2}\omega L$ and 
$\pm$ refer to seed photon polarisation being perpendicular or parallel to 
that of 
the external field and we have adapted the rate for photon-splitting from 
\cite{papanyan72}. Photon-splitting 
requires dispersion to be taken into account and has a strong dependence on the 
frequency being split $\sim (\omega/m)^{5}$, whereas photon merging requires a 
high density of photons such that the number of seeds in a cylindrical volume 
of radius $\lambdabar$ around the photon's trajectory is not too small. 
Although a full comparison is beyond the scope of this paper, if one notes that 
in a photon gas at temperature $T$ the density of photons with energies 
$\in [\omega, \omega+\delta \omega]$, $\delta \omega/\omega \ll 1$ is of the 
order $\rho\sim\omega^{2}\delta\omega[\exp(\omega/T)-1]^{-1}$, then the ratio 
of second harmonic generation to photon splitting is of the dependency
\bea
\frac{\rho_{2\gamma\to\gamma'}}{\rho_{\gamma\to2\gamma'}} &\sim& 
\frac{L\,\delta\omega}{\mbox{e}^{\omega/T}-1}\\
\frac{\widetilde{\rho}_{2\gamma\to\gamma'}}{\rho_{\gamma\to2\gamma'}} &\sim& 
\left(\frac{m}{\omega}\frac{\Delta 
}{B^{2}}\right)^{2} 
\frac{\lambdabar}{L}\frac{\delta \omega}{m} \frac{1}{\mbox{e}^{\omega/T}-1}
\eea
When is harmonic generation more prevalent than 
electron-positron pair 
creation in a strongly-magnetised thermal photon gas? If the number density of 
pairs created in photon-photon collisions is $\rho_{2\gamma\to e^{+}e^{-}}$ and pairs created 
through photon decay in a background constant magnetic field $\rho_{\gamma\to e^{+}e^{-}}$ 
then
\bea
\rho_{2\gamma\to e^{+}e^{-}} &\sim& 2 
\frac{1}{\lambdabar^{3}}\frac{L}{\lambdabar} 
\left(\frac{\alpha}{2\pi}\right)^{2} 
\left(\frac{T}{m}\right)^{3}\mbox{e}^{-\frac{2m}{T}} \label{eqn:pair1}\\
\rho_{\gamma\to e^{+}e^{-}} &\sim& \frac{3^{3/4}\alpha 
}{4\sqrt{2}\pi^{3/2}}\,\frac{1}{\lambdabar^{3}}\frac{L}{\lambdabar} 
\,\left(\frac{T}{m}\right)^{2}\delta^{1/4} 
\mbox{e}^{-\frac{4}{\sqrt{3\delta}}},  \label{eqn:pair2}
\eea
for $T/m\ll 1$ and $\delta = TB/2m \ll 1$ where the pair-creation 
densities were adapted from \cite{king12b, king12c} for a constant magnetic 
background. In order to calculate the total density of merged photons created 
in a photon gas, we would have to extend our calculation to include merging of 
photons with different 
wavevectors and integrate the double-photon rate over a double Bose-Einstein 
distribution. However, from \eqnrefs{eqn:pair1}{eqn:pair2} we already note that 
for $T/m\ll1$, pair-creation is exponentially suppressed whereas photon merging 
(and splitting) are perturbative in $T/m$. Since 
$T/m \sim 10^{-4}$ for strongly-magnetised neutron stars \cite{harding06}, 
one could pose 
the question whether harmonic generation, along with photon splitting, can be 
an 
important factor in the evolution of these stellar objects.
\newline

We close by noting that only the asymptotic photon merging signal is of 
relevance to laser physics, and then only when the laser background occurs to 
an even power and hence contains a slowly-varying component. This occurs in 
six-photon scattering if 
pulses collide at an angle which is proportional to $\ampp^{3}\amps^{2}$, 
or in 
eight-photon scattering which is proportional to $\ampp^{3}\amps^{4}$ and 
considering 
that $\mathcal{E}_{p,s} \ll 1$, these 
signals are greatly suppressed. This suppression can be potentially overcome 
by 
using an ultra-short strong laser pulse and looking off-axis for emitted 
photons \cite{king12} using more than two laser 
frequencies and off-axis 
beams \cite{lundstroem_PRL_06}, or using a charged projectile such as a proton 
\cite{dipiazza08,dipiazza08b}.
\newline
%%%%%%%%%%%%%%%%%%%%%%%%%%%%%%%%%%%%%%%%%%%%%%%%%%%%%%%%%%%

%%%%%%%%%%%%%%%%%%%%%%%%%%%%%%%%%%%%%%%%%%%%%%%%%%%%%%%%%%%
\section{Conclusion}
We have shown that when an oscillating probe field propagates through a 
background field 
with some inhomogeneity, a source of photon-photon scattering appears when the 
two fields overlap \colr{and the field invariants are non-vanishing}. This ``overlap field'' 
disappears when the overlap of the 
two fields tends to zero and is distinct from the ``asymptotic 
field'' that persists after scattering has taken place. Moreover, the overlap 
field permits high harmonic generation for a specific harmonic at an order of 
the fine structure constant lower than in the asymptotic field. By 
integrating the weak-field expansion of the Heisenberg-Euler Lagrangian 
using the Green's function for the wave equation in one spatial and one 
temporal dimension, we compared the nature of the overlap and asymptotic fields 
and identified a suitable non-linearity parameter. We have 
highlighted the potential importance of this effect in astrophysical 
environments by calculating the density of merged photons and contrasted this 
with the density of photons split and density of photons seeding pair creation.
%%%%%%%%%%%%%%%%%%%%%%%%%%%%%%%%%%%%%%%%%%%%%%%%%%%%%%%%%%%

\section{Acknowledgments}
B. K. acknowledges the hospitality of H. R. and the Arnold Sommerfeld Center 
for Theoretical Physics at the Ludwig Maximilians University as well as useful 
editorial suggestions from T. Heinzl. P. B. acknowledges 
the very useful advice of A.
Hindmarsh during development of the computational simulation. This work was 
supported by 
Grant No. DFG, FOR1048, RU633/1-1, by SFB TR18 project B12 and by the 
Cluster-of-Excellence ``Munich-Centre for Advanced Photonics'' (MAP). Plots
were generated with {\tt{Matplotlib}}
\cite{matplotlib}.

\bibliography{current}
\end{document}